\documentclass[journal,comsoc]{IEEEtran}
\usepackage[table]{xcolor}

\usepackage[linesnumbered,ruled,vlined]{algorithm2e}

\usepackage[T1]{fontenc}

\usepackage[table]{xcolor}
\usepackage{tabularx}
\usepackage{cellspace}
\usepackage{graphicx}
\usepackage{amssymb}

\usepackage{newtxmath}

\ifCLASSINFOpdf
\else
\fi

\usepackage{bm}
\usepackage[cmintegrals]{newtxmath}

\usepackage{balance}
\usepackage[caption=false]{subfig}
\usepackage{graphicx}
\usepackage{subfig}
\usepackage{verbatim}
\usepackage{array}
\usepackage{multirow}
\usepackage{ragged2e}
\usepackage{booktabs}
\usepackage{subfig}
\usepackage{float}
\usepackage{url}
\usepackage{tabularx}
\usepackage{tikz}
\usepackage[hidelinks]{hyperref}
\usepackage{adjustbox}
\usepackage{booktabs}
\usepackage{rotating}
\usepackage{slashbox}
\usepackage{xcolor}
\usepackage{color}
\usepackage{colortbl}
\usepackage{multirow}
\usepackage[misc,geometry]{ifsym}
\usepackage{dirtytalk}
\usepackage[normalem]{ulem}

\usepackage{svg}
\usepackage{amsmath} 
\usepackage{physics} 
\usepackage{blindtext}
\usepackage[ruled,vlined]{algorithm2e}
\definecolor{grey}{RGB}{197,197,197}
\newcommand{\orcid}[1]{\href{https://orcid.org/#1}{\includesvg[width=10pt]{orcid}}}
\definecolor{grey}{gray}{0.9}
\usepackage{cite}
\usepackage{lineno}
\usepackage{fancyhdr,graphicx,amsmath,amssymb}
\usepackage[bottom]{footmisc}
\colorlet{mygreen}{green!60!gray}

\usepackage{hyperref} 
\usepackage{tablefootnote}
\usepackage{threeparttable}
\usepackage{cellspace} 
\begin{document}

\title{Federated Learning Under Attack: \\ Exposing Vulnerabilities through Data\\ Poisoning Attacks in Computer Networks}

\author{
Ehsan~Nowroozi,~\IEEEmembership{Senior Member,~IEEE,} Imran Haider, ~\IEEEmembership{Member,~IEEE,} Rahim Taheri , ~\IEEEmembership{Member,~IEEE}, Mauro Conti, ~\IEEEmembership{Fellow Member,~IEEE}
\IEEEcompsocitemizethanks{

\IEEEcompsocthanksitem E. Nowroozi is with the Department of Business and Computing, Ravensbourne University London (RUL), London, United Kingdom (e-mail: e.nowroozi@rave.ac.uk) - Corresponding Author is Ehsan Nowroozi.
\IEEEcompsocthanksitem I. Haider is with the Department of Natural Engineering and Sciences, Bahcesehir University (BAU), Istanbul, Turkey (e-mail: imran.haider@bahcesehir.edu.tr)
\IEEEcompsocthanksitem R. Taheri is with the School of Computing, Faculty of Technology, University of Portsmouth, United Kingdom (e-mail: rahim.taheri@port.ac.uk)
\IEEEcompsocthanksitem M. Conti is with the Department of Mathematics, Security and
Privacy Research Group, University of Padua, 35121 Padua, Italy, and also with the Faculty of Electrical Engineering, Mathematics and Computer Science, Delft University of Technology, 2628 CD Delft, The Netherlands (e-mail: mauro.conti@unipd.it).
}
}

\maketitle

\begin{abstract}
Federated Learning (FL) is a machine learning (ML) approach that enables multiple decentralized devices or edge servers to collaboratively train a shared model without exchanging raw data. During the training and sharing of model updates between clients and servers, data and models are susceptible to different data-poisoning attacks.

In this study, our motivation is to explore the severity of data poisoning attacks in the computer network domain because they are easy to implement but difficult to detect. We considered two types of data-poisoning attacks, label flipping (LF) and feature poisoning (FP), and applied them with a novel approach. In LF, we randomly flipped the labels of benign data and trained the model on the manipulated data. For FP, we randomly manipulated the highly contributing features determined using the Random Forest algorithm. 
The datasets used in this experiment were CIC \cite{Leevy2020} and UNSW \cite{7348942} related to computer networks. We generated adversarial samples using the two attacks mentioned above, which were applied to a small percentage of datasets. Subsequently, we trained and tested the accuracy of the model on adversarial datasets. We recorded the results for both benign and manipulated datasets and observed significant differences between the accuracy of the models on different datasets. From the experimental results, it is evident that the LF attack failed, whereas the FP attack showed effective results, which proved its significance in fooling a server. With a 1\% LF attack on the CIC, the accuracy was approximately 0.0428 and the ASR was 0.9564; hence, the attack is easily detectable, while with a 1\% FP attack, the accuracy and ASR were both approximately 0.9600, hence, FP attacks are difficult to detect. We repeated the experiment with different poisoning percentages.

\end{abstract}


\section{Introduction}

\IEEEPARstart{I}{n} this modern era of technology, in which other scientific disciplines are advancing swiftly, FL is also keeping pace and progressing rapidly. In the FL setting, a central server collects information from different connected clients regarding their data in a secure manner, such as gradient updates. In this setup, instead of sharing data with a central server, only an updated model is sent or received by a client in the form of gradients or model parameters \cite{shejwalkar2021drawing}. Sending data over a network causes data privacy issues and communication overheads. To overcome these challenges, FL techniques allow gradient information to be sent instead of raw data \cite{carlini2019secret}. 
However, sending model updates over a network can also unknowingly leak sensitive information to a central server or any other third party \cite{mcmahan2018learning}. FL has many practical applications and is used in various distributed systems, such as Google's board \cite{hard2019federated}, to predict the next word while typing. In addition, FL potential applications include sentiment assessment, context-driven points, responding to pedestrian actions in self-driving cars, and foreseeing health incidents such as heart attack susceptibility through wearable gadgets \cite{huang2020loadaboost}. 
The wide adaptability of FL across multiple fields has significantly amplified its allure from the perspective of potential threats. While conducting attacks against FL, different threat models are considered, including the objective of the adversary, knowledge of the adversary, and capabilities of the adversary \cite{liu2022threats}. The objective of an adversary is to violate the integrity and availability of a model. The adversary can have complete knowledge of the global model, also known as the white-box setting, in which the adversary can access the data of all collaborating clients, the global model parameters, and its predictions. However, in a black-box setting, the attacker does not have any information regarding the data or model parameters \cite{Privacy_Robustness}. In another case, the adversary can have partial knowledge of the FL setting, which means that the attacker can have access only to the local data of compromised clients but not of benign clients\cite{shejwalkar2021drawing}.

In general, attacks in FL can be divided into two categories (i) causative attacks \cite{8170807}, and (ii) exploratory attacks \cite{luo2023exploreadv}. Causative attacks affect the learning process by controlling training data. These attacks include the injection of training data with erroneous labels, which can reduce the accuracy of the trained model. Typically, these attacks occur prior to model training \cite{inproceedings}. However, exploratory attacks capitalize on misclassifications but do not influence the training procedure \cite{Barreno2010}. There are different kinds of causative attacks that can be carried out against FL such as data poisoning attacks \cite{csoonline_data_poisoning}, model poisoning attacks \cite{cao2022mpaf}, membership inference attacks \cite{shokri2017membership}, 
and many more. These attacks can be classified into various subcategories. There are two types of membership inference attacks: (i) active and (ii) passive. In an active membership inference attack, the attacker could be one of the participants in federated learning who adversarially modifies his parameter uploads $W_t^i$ or could be the central aggregator who adversarially modifies the aggregate parameters $W_t$ that he sends to the target participant(s). However, in a passive membership inference attack, an attacker can only observe genuine computations using the training algorithm and model \cite{Nasr_2019}. Attackers can use both approaches to compromise the clients and servers.\cite{shokri2017membership}.

In this study, our main focus was on data poisoning attacks. 
We implemented two instances of data-poisoning attacks: LF and FP. In LF attacks, the labels of some proportion of the training dataset are flipped so that the attack remains undetectable by humans; for example, if it is 0, then it is changed to 1, and vice versa. By contrast, in FP, we can manipulate some values of the most important columns of the data, and we can use different techniques for it, such as replacing them with the mean or mode of that column.

Most importantly, we determined that the LF attack is not effective in FL, specifically in Computer Networks. The reason for this is the drastic drop in the server accuracy from 90\% to approximately 10\% after applying the LF attack, making it easy for the system to understand that there is an attack. Moreover, the ASR also increased to almost 90\%, which makes the attack more suspicious, and we cannot fool the system by applying this attack to the client because the system can easily detect it. On the other hand, the FP attack was successful in fooling the system because there was no major downward transition in the accuracy of the server, whereas the ASR exhibited a radical and significant upward shift, and this attack remained undetectable to the system, making it highly dangerous.

\subsection{Contributions}
The following points highlight the contributions of this study:
\begin{itemize}
  \item In this study, we conducted training sessions for neural networks using two prominent datasets in the field of computer networks, namely the CIC and UNSW datasets. Our choice of these datasets was driven by their widespread use and relevance in the domain. The primary objective of our neural network training was to perform binary classification of network traffic, distinguishing between benign (0) and malicious (1) traffic. Through these efforts, we aimed to contribute valuable insights and advancements to the understanding and detection of malicious activities in computer networks.
  
  \item In the FL configuration employed for this study, our experimental setup comprised two clients and a server. Specifically, in a white-box scenario, we designed and implemented two data-poisoning attacks, namely label flipping (LF) and feature poisoning (FP), exclusively on one of the clients. The objective was to assess the repercussions of these attacks on the accuracy of the server. Our investigation into these targeted adversarial techniques contributes valuable insights into the vulnerabilities and potential mitigations within the FL framework.
  
  \item Our study delves into the efficacy of data poisoning attacks within the computer network domain, adopting an innovative approach by implementing these attacks at varying percentages on the training data of a single client. This experimentation, conducted on datasets specifically curated for the computer network domain, enabled us to assess the impact of such attacks comprehensively. We calculated both server accuracy and Attack Success Rate (ASR) across different attack percentages, providing a nuanced understanding of the vulnerabilities introduced by data poisoning. To facilitate the reproducibility and adaptability of our experiments, we developed a highly flexible and generic codebase. This codebase allows for seamless adjustment to diverse datasets and facilitates experimentation with different attack percentages. Our contribution lies not only in the insights gleaned from our experiments but also in the provision of a tool that can be easily employed and extended for future research in this area.

  \item In our investigation, we recorded the server accuracy and Attack Success Rate (ASR) across varied poisoning scenarios for both datasets. Our findings underscore the effectiveness of data poisoning attacks when implemented at different percentages, revealing their impact on the accuracy of the server. Additionally, we provide a comprehensive analysis of the incurred loss experienced by both clients, shedding light on the broader consequences of these attacks. This contribution elucidates the nuanced dynamics of data poisoning in our experimental setting, offering valuable insights into the resilience and vulnerabilities of the system under examination.
  
  
\end{itemize}

\subsection{Organization}
We outline the rest of our paper as follows: In Section ~\ref{Related}, we provide an overview of related works that discuss data-poisoning attacks using different datasets. In Section \ref{sec:methodology}, we discuss the datasets and the network architecture of our BAU1 model. This section also includes the experimental setup and techniques that we applied to perform both data-poisoning attacks.
Section~\ref{Results} presents the results of our experiments performed under different scenarios. In Section ~\ref{Conclusions} we summarize our study and propose future work in the field of FL. 

\section{Related Works} \label{Related}

In data poisoning, different types of attacks are used by researchers and adversaries to manipulate data, such as two important data-poisoning attacks, LF and FP. LF attacks have extensive applications in image processing, as in \cite{paudice2018label} authors evaluated the performance of LF attacks and proposed a defense mechanism on two real datasets from the UCI repository \cite{misc_uci_ml_repository}: MNIST\cite{deng2012mnist} and Spambase\cite{misc_spambase_94}, which are common benchmarks for classification tasks. Furthermore, in \cite{muñozgonzález2017poisoning} the authors used an LF attack to manipulate the MNIST dataset and trained a CNN on a benign and label-poisoned dataset. They found that the attack slightly increased the classification error after injecting ten poisoning points into the training data. The experiment was repeated using the same setting. They used a Multiclass Logistic Regression classifier instead of a CNN and found that the error increased from 2\% to 2.1\% after a random LF attack. The approach outlined in \cite{feng2019learning} can be readily expanded to encompass a label-specific scenario, where the adversary can adjust the predictions of the victim classifiers based on predetermined rules as opposed to merely generating incorrect predictions. Experiments were repeated on various datasets, including CIFAR-10\cite{misc_CIFAR_10} and a reduced version of ImageNet, and the effectiveness of the proposed method was confirmed.

Poisoning attacks are extensively employed for model poisoning and data poisoning purposes, as in \cite{raza2023using} they used different flavors of state-of-the-art data poisoning attacks such as random LF, random label and FP, label swapping, and FP attacks. Moreover, LF attacks are widely used by researchers owing to their simple nature. This attack was prominent in this study  \cite{tolpegin2020data} in which they carried out it on the CIFAR-10 and Fashion-MNIST datasets. They introduced a small percentage of malicious participants, ranging from 2\% to 50\%, that contained malicious training data manipulated using an LF attack. For example, in CIFAR-10 image classification, airplane → bird denotes that the airplane image label is changed to a bird. This attack was intended to boost the likelihood of the global model making incorrect classifications, especially by leading to more frequent misclassifications of airplane images as bird images during testing, and they achieved effective results.

Label poisoning was introduced in this study \cite{HALLAJI2023110384} and was generated through GANs and used to train local models. The datasets used were UNSW \cite{7348942} and NIMS \cite{ALSHAMMARI20111326} which are related to computer networks. Furthermore, they proposed two defense methods to mitigate LF attacks on the FL. In addition to the LF attack, another data-poisoning attack known as the FP attack plays a critical role in FL security. Many datasets have a massive number of features, ranging from hundreds to thousands, and selecting the features to poison is a critical challenge known as feature importance. ML feature selection is also widely used for generalization and has both advantages and disadvantages.

We have advanced subcategories of data-poisoning attacks that can also be used by attackers in FL, such as generative adversarial network (GAN) attacks \cite{yang2017generative}. In this type of attack, the adversary attempts to generate data that closely resembles legitimate data in a collaborative learning environment. However, there are two limitations to this type of attack: the attacker should have background information regarding the victim's data, and the other requirement is that all members of the class are similar \cite{hitaj2017deep}. Furthermore, another advanced type of data poisoning attack aims to produce perturbed samples in the training data using auto-encoders (AE). An auto-encoder (AE) is a type of neural network that tends to recreate its input data, and adversaries can use it to generate malicious samples for training data \cite{feng2019learning}.

In their research \cite{Nguyen2020PoisoningAO} they demonstrated that Intrusion Detection Systems (IDS) for IoT based on FL are prone to backdoor attacks. In their proposed attack strategy, the threat actor(s) can fool the detection model by employing compromised IoT devices to introduce minimal quantities of poisoned data during the training procedure while remaining undetected throughout the process.
Moreover, researchers are actively working on new defense methodologies to reduce data-poisoning attacks. However, more work is required to make FL systems more robust to these attacks, which also evolve. In Table \ref{tab:comparison_related_work} we have compared the work of other researchers with ours and listed the model(s) used, datasets utilized, strengths, and weaknesses.
\begin{table*}
\centering
\renewcommand{\arraystretch}{0.5} 
\caption{Comparison between related and our work}
\resizebox{\linewidth}{!}{%
\begin{tabularx}{\textwidth}{>{\centering\arraybackslash}p{1cm}>{\centering\arraybackslash}p{1.5cm}>{\centering\arraybackslash}p{2cm} X X} 
\toprule
Ref. & Model(s) & Datasets & Strength(s) & Weakness(es) \\
\midrule
\cite{paudice2018label} & K-NN & MNIST \cite{deng2012mnist}  & - Proposed Defense mechanism against LF attack & \RaggedRight - Used a very simple linear classifier \\
& &Breast Cancer\cite{misc_breast_cancer}, SpamBase\cite{misc_spambase_94} & & \RaggedRight - Defense mechanism is not effective for other types of poisoning attacks \\
\midrule
\cite{muñozgonzález2017poisoning} & LR & SpamBase \cite{misc_spambase_94} & - extension of poisoning attacks from binary to multiclass problems.& \RaggedRight - High complexity of Back-Gradient Optimization \\
& MLP &Ransomware\cite{ransomware_automated} & - attack transferability is also possible. &\\ 
\midrule
\cite{feng2019learning} & AE &CIFAR10\cite{misc_CIFAR_10}  & - Transferability Across Classifiers & \RaggedRight - Dependence on Hyperparameters \\
& & ImageNet\cite{deng2009imagenet} & &\\ 
\midrule
\cite{raza2023using} & CNNs & MIT-BIT arrhythmia\cite{MIT_BIT} & - Provide framework to detect poisoning attacks& \RaggedRight - White-box assumption \\
& &HAR\cite{misc_human_activity_recognition} & - low computations complexity & \RaggedRight - Data privacy concerns.\\
\midrule
\cite{HALLAJI2023110384} & GANs & UNSW\cite{7348942} & - Proposed defense against LF attack & \RaggedRight - Scalability and efficiency concerns \\
& &NIMS \cite{ALSHAMMARI20111326} & - combined generative adversarial schemes with noisy-label classifiers & \RaggedRight -  Adversarial training, GANs, and noisy-label classifiers computationally complex.\\
\midrule
\cite{yang2017generative} & GANs & MNIST \cite{deng2012mnist}  & - Legitimate data generation & \RaggedRight - Limited Validation on Diverse Datasets \\
& &CIFAR-10 \cite{misc_CIFAR_10} & - Generative Method for Accelerated Attack Generation & \RaggedRight - Background information on the victim's data is required.\\
\midrule
Ours & NNs & UNSW \cite{7348942}  & - LF and FP attack on two computer networks datasets & \RaggedRight - Effective in white-box scenario \\
& &CIC \cite{Leevy2020} &   & \RaggedRight - Lack of discussion on defense mechanisms \\ 
\bottomrule
\label{tab:comparison_related_work}
\end{tabularx}
}
\label{tab:Table1}
\end{table*}

\section{Methodology} \label{sec:methodology}

\begin{figure}[h!]
    \centering
    \includegraphics[width=0.40\textwidth,height=4cm]{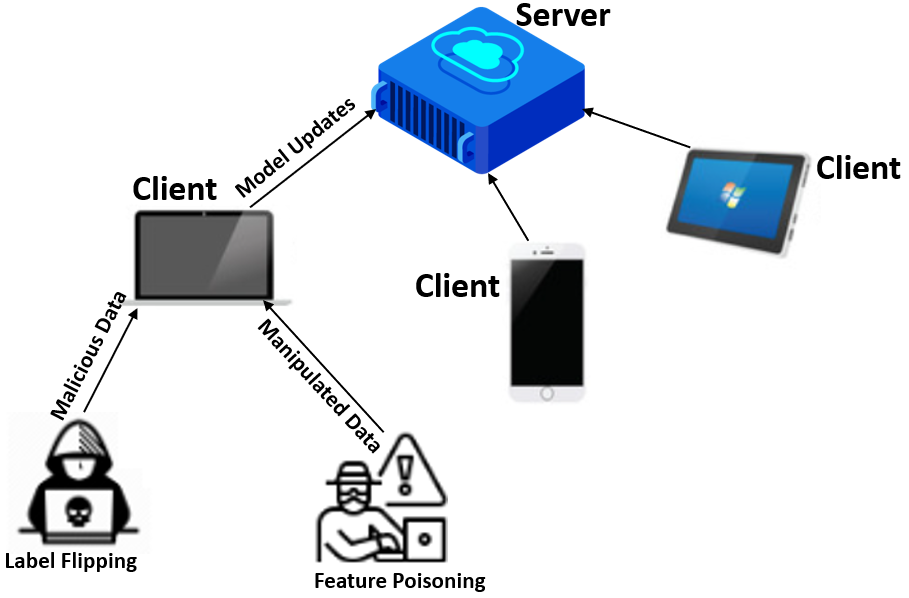}
    
    \caption{Illustration of Adversarial Data Manipulation: A Server and Two Clients. Trained CL1 on Manipulated Data with LF and FP Attacks}
    \label{fig:Clients_server_architecture}
\end{figure}

In this study, we used the two most popular datasets, CIC \cite{Leevy2020} and UNSW \cite{7348942} related to computer networks. We trained two DL models on these two datasets separately and captured their accuracy. In this FL setup, we considered just two clients and a server and then also applied two data poisoning attacks, LF and FP to only one client. In DL, an attacker can modify the training dataset based on knowledge and information. According to the currently existing literature, there exist three scenarios that an adversary can consider to conduct adversarial attacks: white-box, gray-box, and black-box attack scenarios\cite{Nasr_2019}. The adversary possesses perfect knowledge (PK) about the training data and model in a \textit{white-box} setting and can generate adversarial instances for training and modify the model updates. In the case of \textit{a gray-box} scenario, the threat actor has limited knowledge (LK) of the training data and model. The adversary does not know the internal information of the system in a \textit{black-box} setup, which is a more viable and complex case than the other scenarios. Consequently, the attacker employs recurring inquiries to gather such sensitive data. We carried out this experiment in a \textit{white-box} scenario as we had access to both data and model. Additionally, a very simple illustration of our methodology is given in Figure.\ref{fig:Clients_server_architecture}.
Detailed information about the datasets and data poisoning attacks is provided below:
\subsection{Datasets}
To examine the effect of adversary capability on models by data poisoning in computer networks, we used the most popular datasets, the CIC\cite{Leevy2020} and the UNSW \cite{7348942} that are part of many research works \cite{HALLAJI2023110384, nowroozi2022resisting, nguyen2023robust}. We used them to train the model on both benign and malicious examples. Therefore, we used two datasets that are highly related to our research domain. We gathered these datasets in the form of pcap files and then extracted those pcap files using NFStream \cite{nfstream}. NFStream employs nDPI-based deep packet inspection to identify encrypted applications and perform metadata fingerprinting accurately. This includes the detection of protocols such as TLS, SSH, DHCP, and HTTP. nDPI stands for "nTop Deep Packet Inspection." It is an open-source library and toolkit used for deep packet network traffic inspection. Deep packet inspection involves analyzing the content of network packets at a granular level to understand the protocols, applications, and services being used on a network. nDPI is commonly used in network monitoring and security applications to classify and analyze network traffic, making it easier to detect and respond to security threats, optimize network performance, and gain insights into network usage patterns.

We trained the clients on both CIC and UNSW and in all scenarios,  divided the dataset in this way: 80\% of the data for the training and 20\% of data for testing and validation. Moreover, each dataset contains around one million data samples. Furthermore, partitioned the training dataset into two segments, allocating one segment to CL1 and the other to CL2. Additionally, subdivided 20\% of testing data into two separate sets, distributing 10\% for validation and reserving the remaining 10\% for testing. 
Before giving data to the model, also pre-processed it for efficient results such as filling the missing values with a specified number. If the number is not specified then those values are filled with 0. In addition, during the pre-processing of the data normalization was applied because it gives improved convergence, enhanced model performance, equalized influence, and interpretability because normalized data has consistent scales, that simplify the comparison of feature importance.

\subsection{Network Architecture}
\begin{figure}[h!]
    \centering
    \includegraphics[width=0.49\textwidth,height=3.3cm]{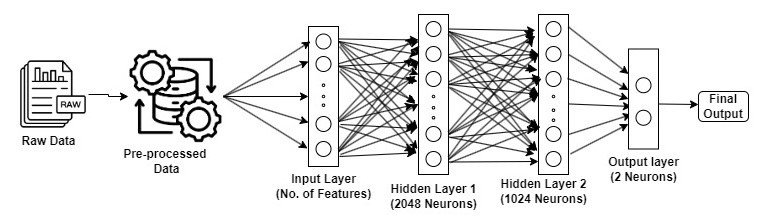}
    
    \caption{Architecture of Neural Network Model (We referred this network to as BAU1)}
    \label{fig:network_architecture}
\end{figure}

We adopted a single deep-learning model for both clients during the training and testing phases. Initially, this same model was utilized for our server as well. This model was composed of an input layer, and the number of neurons in this layer is set to the feature size, where feature size represents the number of (Total\_Columns-1) in the dataset, and we subtracted one column because one column represents the output label column and is not part of the feature columns. Its number of output neurons was 2048. The model also contains two hidden layers, one of which contains 2048 input neurons, whereas the number of output neurons is 1024, which is the number of input neurons for the next hidden layer. Similarly, there are two output neurons in the second hidden layer, which is equal to the number of output classes. Moreover, the model was also constructed from two activation functions and used the Rectified Linear Unit (ReLU) activation function named ReLU1, which was applied after the first hidden layer, and ReLU2, which was applied after the second hidden layer. The architecture of the neural network model, named BAU1, can be easily obtained from Figure \ref{fig:network_architecture}. 
Furthermore, batch normalization was applied to both hidden layers. Normalization is performed during pre-processing and can lead to more stable and effective ML models. A dropout layer with a specific dropout probability ($\text{self.dropout\_p}$) is also used as part of the neural network architecture. During training, it randomly deactivates some neurons to increase the network's proficiency for generalization and diminish overfitting. The output activation function that we used was the log softmax function, which computes the natural logarithm (base e) of softmax probabilities. The cross-entropy loss function is used in this network, which is commonly used for classification problems to compute the loss between the predicted probability distribution and the labels of the true class. The output layer of our model comprises two neurons, and each neuron represents one of the binary classes (0 or 1), where 0 represents the benign class and 1 corresponds to the malicious class.

\subsection{Experimental Setup} 
To build the model, we considered 8,38,861 samples for training, 1,04,857 for validation, and 1,04,857 for the test set. In Table \ref{tab: Table1 } different attack scenarios are presented, in which $N_{BAU1-LF}^{UNSW}$ represents the NN model named BAU1, which is trained with the UNSW dataset, and to this dataset, the LF attack was applied. Similarly, this scenario, $N_{BAU1-FP}^{CIC}$, indicates that we trained BAU1 with the CIC dataset that is poisoned using the FP attack. In addition, another important scenario $N_{BAU1-LF}^{CIC}$ points to the LF attack on the CIC dataset, which is used to train the BAU1 model, whereas $N_{BAU1-FP}^{UNSW}$ denotes the FP attack on the UNSW dataset that is fed to the BAU1 model for training. The number of samples in both datasets is also equal, that is, 10,48,575 examples in each dataset. Most DL models are designed to accept images as input, which are usually three-dimensional, and the dataset we used was one-dimensional. Therefore, to make the model feasible, two additional dimensions were added. After adding extra dimensions, the shape of the np array was changed, and it was performed for both clients' training data and testing and validation data. We developed an NN using PyTorch \cite{pytorch-docs} which is a popular Python library. We performed our analysis using hardware specification MSI GF65. We have made all our implementation codes publicly available in the GitHub repository \cite{Github_FL}. For the NN model, we considered 20 epochs for training using the Stochastic Gradient Descent (SGD) optimizer with a random learning rate between $1 \times 10^{-4}$ and $9.9 \times 10^{-3}$ and a momentum of 0.9. The batch size for training and validation is set to 1000. For both clients, we used the Federated Averaging (FedAvg) algorithm \cite{FedAvg} which is a common setup in FL, where multiple clients with local datasets train models in a collaborative manner while preserving data privacy.

\subsection{Empirical Study}
In this experiment, we performed two data-poisoning attacks on two models that were trained on two different datasets. First, an LF attack was applied. Before performing this attack, we trained the model with a benign dataset and captured the results, saved the models, and tested the accuracy of the saved models using test data. We then poisoned the data by flipping the labels for 1\% of the data. Again, the model was trained with malicious data, and the results and models were saved. Here, to compute the \textit{ASR}, we flipped the labels of the complete test data and noted the accuracy that indicates the strength of the attack. We repeated the experiment with 2\%, 3\%, 4\%, 5\%, 7\%, 10\%, 15\%, 20\%, and 25\% poisoned data, documented the results for all these attack percentages, and computed the ASR for all of them.

\begin{figure}[h!]
    \centering
    \includegraphics[width=0.50\textwidth,height=6cm]{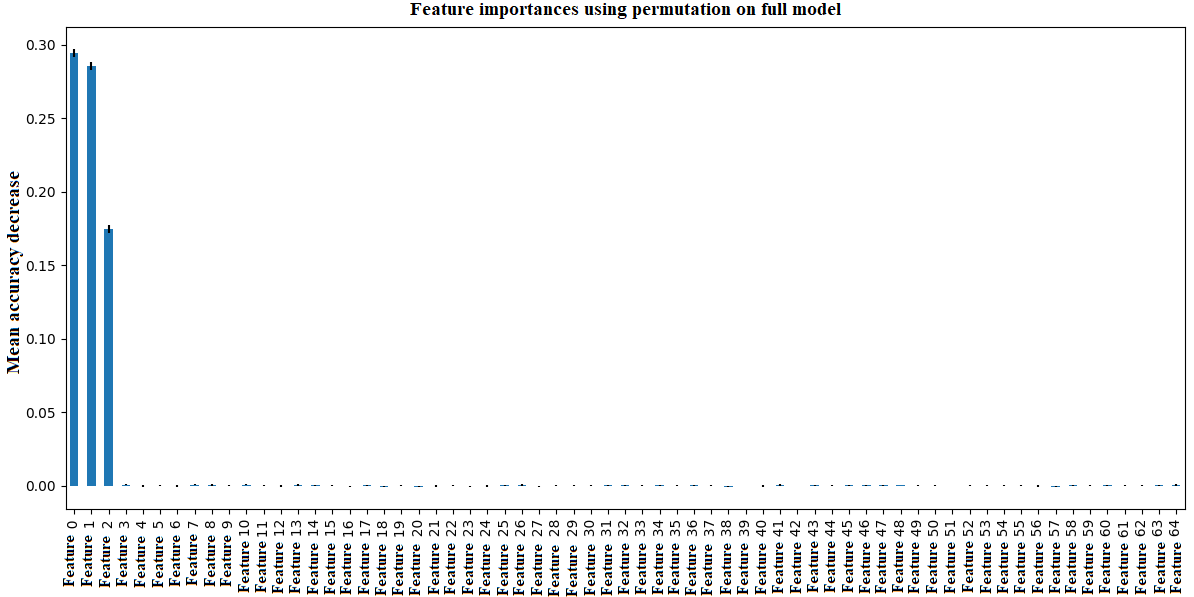}

    \caption{CIC: Feature importance using Permutation on the full model}
    \label{fig:permutation-cic}
    
\end{figure}

\begin{figure}[h!]
    \centering
    \includegraphics[width=0.49\textwidth,height=6cm]{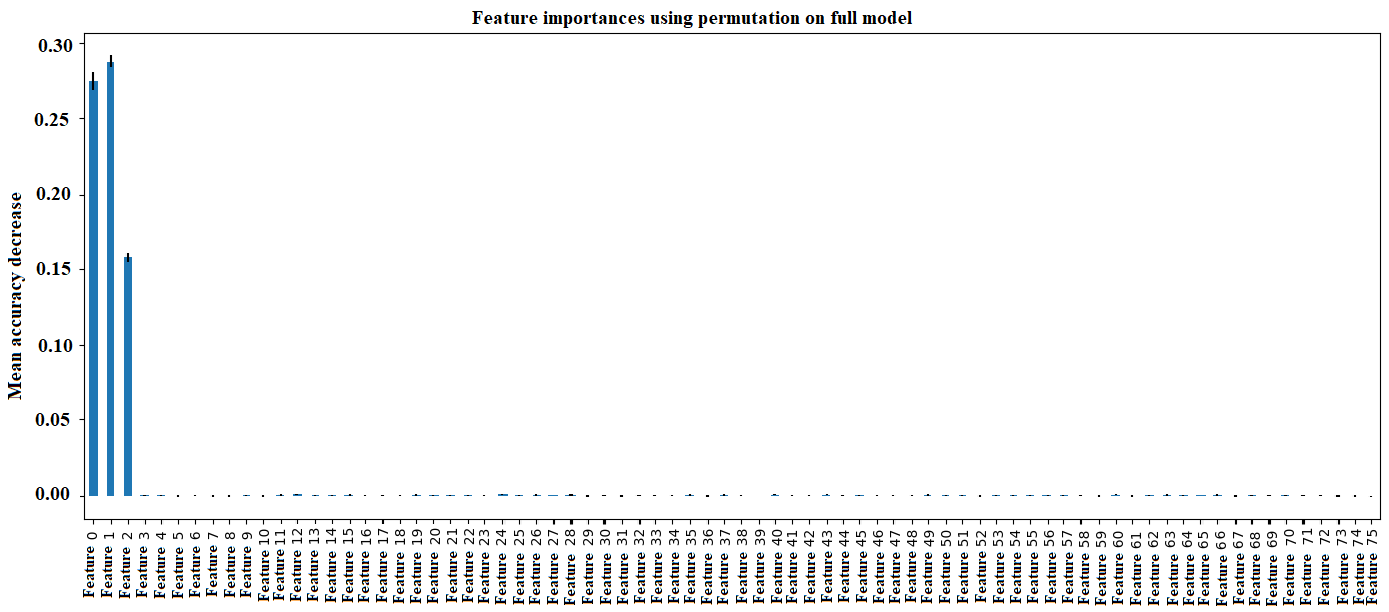}
    
    \caption{ UNSW: Feature importance using Permutation on the full model}
    \label{fig:permutation-unsw}
\end{figure}

We performed the same experiment for both the CIC and UNSW datasets. Our code is written in such a generic manner that simply changes the name of the dataset; the rest of the processes, for example, pre-processing of data, splitting the training data for both clients, training of the model with/without attack, computation of ASR, and saving the results in an Excel file, are performed for both datasets without any user interaction. For the FP attack, we introduced some new changes in the code; however, we used a random forest algorithm \cite{scikit-learn_forest_importances} to compute feature importance. In the CIC dataset, the \textit{first column} has the highest feature importance, which we computed using permutation on the full model, as shown in Figure. \ref{fig:permutation-cic}. Permutation-based feature importance is an approach that involves logically rearranging the values of each feature, analyzing the influence on the performance of the model, and hence determining the impact intensity of each feature in the decision-making step.
We used a similar strategy to determine the feature importance in the UNSW dataset and found that \textit{second column} has the highest feature importance. The graph of all columns with their feature importance is given below for this particular dataset in Figure \ref{fig:permutation-unsw}.
In FP, we manipulated the values of the most important features that were determined using the random forest technique. There were two label classes in both datasets: 0 and 1. We first calculated the \textit{mean} of all values where the label is 0 and computed the mean of values with label 1. In the next step, we found the minimum and maximum values in the feature column that are different for the two datasets. Subsequently, the values of the most important features were normalized using the \textit{min-max} normalization technique. We changed the values of the column where the label was 0 to random unique values, where the label was 1. Detailed information regarding our FP attack can be found in Algorithm \ref{algo:FP_Attack}. In this algorithm, $D$ represents a dataset that contains feature column values. \textit{att\_per} represents the attack percentage which is also represented as P in the algorithm, which is the percentage of values that are to be manipulated. $\text{min\_value}$ and $\text{max\_value}$ represent the minimum and maximum values in the feature column, respectively. The \textit{i} represents the iterator value of the loop, L represents the value of the label that may be either 0 or 1, and \textit{percent} represents the number of values to be manipulated in the feature column. Additionally, $\text{number}$ represents the user-defined att\_per, and $\text{column\_index}$ is the index of the target feature column.
\begin{table}[h!]
    \centering
    \renewcommand{\arraystretch}{1.5} 
    \caption{Attack Scenarios on BAU1 Network with different Datasets}
    \begin{tabular}{|c|c|c|} \hline 
         Scenarios&  Dataset & Attack Strategy\\ \hline 
         $N_{BAU1-LF}^{UNSW}$&  UNSW & Label Flipping\\ \hline 
         $N_{BAU1-FP}^{UNSW}$&  UNSW & Feature Poisoning\\ \hline 
         $N_{BAU1-LF}^{CIC}$&  CIC & Label Flipping\\ \hline 
         $N_{BAU1-FP}^{CIC}$&  CIC & Feature Poisoning\\ \hline
    \end{tabular}
    \label{tab: Table1 }
\end{table}

\begin{algorithm}
    \caption{Feature Poisoning (FP) Attack}
    \label{algo:FP_Attack}
    \KwData{%
        $D$: Feature column values,\\
        $L$: Labels (0 and 1),\\
        $\text{att\_per}$: P
    }
    \KwResult{Transformed feature values}
    
    \textbf{Step 1:} Find the min and max values in feature column\;
    \hspace{1cm} $min\_value \leftarrow \min(D)$\;
    \hspace{1cm} $max\_value \leftarrow \max(D)$\;
    
    \textbf{Step 2:} find the average for label classes 0 and 1\;
    \hspace{1cm} $average\_zero \leftarrow \text{average}(D \text{ where } L = 0)$\;
    \hspace{1cm} $average\_one \leftarrow \text{average}(D \text{ where } L = 1)$\;
    
    \textbf{Step 3:} Normalize the values\;
    \For{i in dataset}{
        \eIf{$L = 0$}{
            \hspace{1cm} $normalized\_value \leftarrow \dfrac{average\_zero - min\_value}{max\_value - min\_value}$\;
        }{
            \hspace{1cm} $normalized\_value \leftarrow \dfrac{average\_one - min\_value}{max\_value - min\_value}$\;
        }
        \hspace{1cm} Update the feature value with $normalized\_value$\;
    }
    
    \textbf{Step 4:} Modify malicious samples to benign samples\;
    \hspace{1cm} $\text{number} \leftarrow \text{P}$ \;
    \hspace{1cm} $percent \leftarrow \text{int}(\text{len}(L) \times (\text{number} / 100))$\;
    \For{$i = 1$ \KwTo $percent$}{ 
        \hspace{1cm} $random\_index \leftarrow \text{random}(0, \text{len}(unique\_values) - 1)$\;
        \hspace{1cm} \If{$L[i] = 1$}{
            \hspace{1cm} $D[i, 0, 0, \text{column\_index}] \leftarrow unique\_values[random\_index]$\;
        }
    }
\end{algorithm}

In contrast, to calculate the ASR, the value of the column where the label was 0 was replaced with the normalized average value of label 1 and vice versa. Because we considered two clients in this experimental setup, we performed both attacks on only one client, which we named CL1. The percentage of attacks was calculated as the number of values to be manipulated according to the desired percentage value. We set up an array containing integer values. In every iteration, if the iterator value is not in that array, then the iterator value is considered a percentage number. This percentage is multiplied by the number of labels in the training data of CL1, which is also equal to the number of samples/rows in the training data. In equation form, it can be written as:

\begin{equation}
\text{number\_of\_values} = \left\lfloor \frac{\text{len}(CL1\_Y) \cdot \text{attack\_percentage}}{100} \right\rfloor
\end{equation}

In the above equation, CL1\_Y represents the length of the array of labels of training data for CL1, where \textit{len} in Python is used to determine the length of an array, while \_of\_attack represents the percentage number, and the number\_of\_values denotes the number of values that we have to change in the feature column. During computations in Python, this equation can return the float value, so that it changes to the integer value we have used the floor function.

\section{Results and Discussion}
\label{Results}
In this section, we discuss the results obtained during the experiment using different scenarios, with and without an attack.

\subsection{FL model in the presence of NO attack}
\label{subsec:results_without_attack}
We trained the BAU1 model using benign datasets. First, we trained on the CIC dataset without an attack, and reported the results, which are represented as $N_{BAU1}^{CIC}$. We used this notation in Table \ref{tab:without_attack} to clarify the results. Subsequently, we trained the BAU1 model with the UNSW dataset without an attack, which is denoted as $N_{BAU1}^{UNSW}$ in Table \ref{tab:without_attack}. We captured the server accuracy for both the scenarios. Moreover, we had two clients during this experimental study and recorded their losses, as listed in Table \ref{tab:without_attack}. 
\begin{table}[h!]
    \centering
     \renewcommand{\arraystretch}{1.5}
    \caption{Results without Attack}\begin{tabular}{|c|c|c|c|} \hline 
         Scenario&  Component& Loss&Accuracy\\ \hline 
         $N_{BAU1}^{CIC}$&  Client-1 \& Client-2&  0.6406& -             \\ \hline 
         \rowcolor{gray!30}
 $N_{BAU1}^{CIC}$& \textbf{Server}& -&0.9680\\\hline
 $N_{BAU1}^{UNSW}$& Client-1 \& Client-2& 0.6787&-\\\hline
 \rowcolor{gray!30}
 $N_{BAU1}^{UNSW}$& \textbf{Server}& -&0.8027\\\hline
    \end{tabular}
    
    \label{tab:without_attack} 
\end{table}
In this table, we report the loss of client-1 and client-2 and save the values obtained after the last epoch. The loss value for both clients is 0.6406 for the CIC dataset. In the same scenario, the server accuracy was 96.8

\subsection{Results With Label Flipping  (LF) Attack on CIC dataset}
Table \ref{tab:with_LF_attack_CIC} lists the results of the LF attack on the CIC dataset. Before, explaining the results of this scenario, there are some common things in all the result tables in all attack scenarios. We set a threshold value for the successful results, which is 0.40, which means that if the server accuracy and ASR are equal to or greater than 0.40, then the attack percentage will be considered a successful case.

In Table \ref{tab:with_LF_attack_CIC} it is clear that the server accuracy drops significantly and the ASR increases; therefore, an LF attack creates a situation of suspiciousness in the system. The losses for client-1 and client-2 were 0.7352 and 0.7361, respectively. With 1\% data manipulation, the accuracy of the server was transcribed as 0.0428, whereas the ASR is 0.9564, indicating that the attack failed to fool the server because the accuracy of the server dropped drastically to a very small value and the system could understand it very easily. A similar behavior can be observed with a 2\% attack with an accuracy of 0.0537, whereas the ASR is 0.9457, which is also a failure scenario for the LF attack because the attack can no longer act as a ghost in the system and can be detected easily by the system. To further investigate the impact of the LF attack, we increased the attack percentage. We observed new behavior in server accuracy and ASR with 3 \% and 4 \% attacks. The accuracy of the server with 3\% attacks increased to 0.9680, whereas the ASR value dropped to 0.0329, and almost similar results were obtained with a 4\% attack. The increase in accuracy is very high, while the ASR value is very low because ASR is not satisfactory in both 3\% and 4\% of attack cases, and because of the very small ASR values, both cases are considered to fail.

An attack strength of 5\% yields interesting results. The accuracy dropped to 0.7739, whereas the ASR increased to 0.2292, which is not a successful scenario according to the threshold. However, detecting the LF attack with this percentage is difficult because the accuracy has not dropped too low, and the ASR showed some improvement. To simplify our experiment, we do not calculate the results for 6\%, 8\%, or 9\%. With 7\%, 10\%, 15\%, and  20\% of attacks, we observed similar results, where the accuracy of the server was very low, while ASR had a significantly high value, indicating that these failed percentages of the attack. With 25\%, we again observed the opposite behavior, in which the accuracy of the server increased to 0.9204, while ASR decreased to 0.0797; hence, the attack failed with this percentage of attack because of a very small value of ASR. 
\begin{table}[h!]
    \centering
    \renewcommand{\arraystretch}{1.2}
    \caption{LF Poison Attack for the Scenario $N_{BAU1}^{CIC-LF}$ on CIC Dataset}
    \begin{tabular}{|c|c|c|c|c|} \hline 
         Poison & Component &Loss & Acc & ASR\\ \hline 
         1\%& Client-1 \& Client-2 &  0.7352 \& 0.7361 & - & -  \\ \hline 
         \rowcolor{gray!30}
         1\%& \textbf{Server} & - &0.0428 &0.9564\\\hline
         
         2\% &Client-1 & 0.7224 \& 0.723 & - & -\\\hline
         \rowcolor{gray!30}
         2\%& \textbf{Server} & - &0.0537 &0.9457\\\hline
         
         3\%& Client-1 \& Client-2 & 0.6378 \& 0.7116 & - & -\\\hline
         \rowcolor{gray!30}
         3\%& \textbf{Server} & - & 0.968 &0.0329\\\hline
         
         4\%& Client-1 \& Client-2 &0.6721 \& 0.6713 & - & -\\\hline 
         \rowcolor{gray!30}
         4\%& \textbf{Server} & - & 0.9486&0.0539\\\hline
         5\%& Client-1 \& Client-2 & 0.6803 \&0.6794 & - & -\\\hline
         \rowcolor{gray!30}
         5\%& \textbf{Server} & - & 0.7739&0.2292\\\hline
        
         7\% &Client-1 \& Client-2 &  0.708 \&0.7093 & - & -\\\hline
         \rowcolor{gray!30}
          7\%& \textbf{Server} & - & 0.1256&0.8720\\\hline
          
         10\%&Client-1 \& Client-2 & 0.7464 \& 0.7534 & - & -\\\hline
         \rowcolor{gray!30}
          10\%& \textbf{Server} & - & 0.032&0.9670\\\hline
          
          15\%&Client-1 \& Client-2 & 0.7129 \&0.7171 & - & -\\\hline
          \rowcolor{gray!30}
          15\%&\textbf{Server}& - & 0.0447&0.9543\\\hline
          
         20\%&Client-1 \& Client-2 &  0.7076 \& 0.7116 & - & -\\\hline
         \rowcolor{gray!30}
          20\%& \textbf{Server}& - & 0.1281&0.8718\\\hline
          
         25\%& Client-1 \& Client-2 & 0.6744 \&0.6669 & - & -\\\hline
         \rowcolor{gray!30}
         25\%&\textbf{Server}& - & 0.9204 & 0.0797\\ \hline
    \end{tabular}
    \label{tab:with_LF_attack_CIC}
\end{table}
\subsection{Results With Label Flipping  (LF) Attack on UNSW dataset} 
This section discusses various cases of LF attacks on the UNSW dataset. We repeated the experiment with one change in data. Previously, we carried out an LF attack on the CIC dataset; however, we changed the dataset to an UNSW, which is also related to computer networks.

In Table \ref{tab:with_LF_attack_UNSW}, starting with a 1\% LF attack, the losses of client-1 and client-2 were 0.6791 and 0.6788, respectively. Furthermore, the accuracy of the server with a 1\% attack rate was 0.8554, whereas the ASR was only 0.1423. This value of ASR is less than the threshold, which is why we did not consider this percentage of the attack as a successful attack. Although the accuracy was high, the ASR was low. We observed more interesting results with the 2\% and 3\% attacks. With 2\% and 3\% LF attacks, the loss of both clients increased to approximately 0.7300, which resulted in a decrease in the server accuracy to approximately 0.1000 and an increase in the ASR  to approximately 0.9000. With a sudden drop in accuracy and an immense increase in the ASR, the system will clearly understand that it is under attack, making this case a failed attack. We want to consider only the attack case as a successful attack that gives high values for both the accuracy of the server and the ASR, because we want our attack to act as a ghost in the system.  The attack with 4\% failed because the ASR value was 0.1917, whereas at 5\%, the ASR value increased to 0.7193 and the server accuracy dropped to 0.2815, which also makes it an unsuccessful attack scenario. Attacks of 7\%, 10\%, and 15\% failed because their ASR values were less than 0.1757 in both 7\% and 10\% and could not cross the threshold value, although the accuracy was greater than 0.8000. In either case, the ASR value is not the desired value. Moreover, the attack was unsuccessful with a 20\% LF attack because the ASR value was 0.8212; however, the server accuracy dropped to 0.1795, making it suspicious in the system. However, with 25\% attack failure, although the server accuracy was logged at 0.7097, the ASR value was 0.2919, resulting in a failed LF attack percentage. Based on these results, we can conclude that the LF attack does not generate effective outputs and fails to fool the server.

\begin{table}[h!]
    \centering
    \renewcommand{\arraystretch}{1.3}
    \caption{LF Poison Attack for the Scenario $N_{BAU1}^{UNSW-LF}$ on UNSW Dataset}
    \begin{tabular}{|c|c|c|c|c|} \hline 
         Poison & Component &Loss & Acc & ASR\\ \hline
         1\%&Client-1 \& Client-2 &  0.6791 \& 0.6788 & - & -  \\ \hline 
         \rowcolor{gray!30}
         1\%& \textbf{Server} & - &0.8554 & 0.1423\\\hline
         
         2\%&Client-1 \& Client-2 &  0.731 \& 0.7317 & - & -  \\ \hline 
         \rowcolor{gray!30}
         2\%& \textbf{Server} & - &0.0951 & 0.9052\\\hline
         
         3\% &Client-1 \& Client-2 & 0.7271 \& 0.7258 & - & -\\\hline
         \rowcolor{gray!30}
         3\%& \textbf{Server} & - &0.1000 &0.8997\\\hline

         4\%& Client-1 \& Client-2 & 0.6734 \& 0.6725 & - & -\\\hline
         \rowcolor{gray!30}
         4\%& \textbf{Server} & - & 0.8072&0.1918\\\hline
         
         5\% &Client-1 \& Client-2 & 0.7041 \& 0.7047 & - & -\\\hline
        \rowcolor{gray!30}
         5\%& \textbf{Server} & - &0.2815 & 0.7193\\\hline
        
         7\% &Client-1 \& Client-2 &  0.6784 \& 0.6771 & - & -\\\hline
          \rowcolor{gray!30}
         7\%& \textbf{Server} & - & 0.8253&0.1757\\\hline

         10\%&Client-1 \& Client-2 & 0.6708 \&  0.6676 & - & -\\\hline
         \rowcolor{gray!30}
          10\%& \textbf{Server} & - & 0.8816 & 0.1181\\\hline
          
          15\%&Client-1 \& Client-2 & 0.6853 \& 0.6835 & - & -\\\hline
          \rowcolor{gray!30}
          15\%&\textbf{Server}& - & 0.6793& 0.3186\\\hline
          
         20\%&Client-1 \& Client-2&  0.7032 \& 0.7059 & - & -\\\hline
         \rowcolor{gray!30}
         20\%& \textbf{Server} & - &0.1795 & 0.8212\\\hline
         
         25\%& Client-1 \& Client-2 & 0.6856 \& 0.6826 & - & -\\\hline
         \rowcolor{gray!30}
         25\%& \textbf{Server} & - & 0.7097 &0.2920\\\hline
    \end{tabular}
    \label{tab:with_LF_attack_UNSW}
\end{table}

\subsection{Results With Feature Poisoning (FP) Attack on CIC dataset}
In Table \ref{tab:with_FP_attack_CIC} there are successful cases that show the effectiveness of the FP attack performed with the CIC dataset on one of the two clients in the FL to fool the server. In this scenario, we considered the poison percentages of the attack as successful, in which both the server accuracy and ASR had values greater than the threshold value, which in this case was set to 0.40. We manipulated the values of one feature of the CIC dataset and trained the model on 1\% manipulated values of the most important feature columns in the dataset. With 1\% poisoned data, we observed that the accuracy of the server was 0.9642 and, more interestingly, the ASR value was 0.9628. The accuracy of the server was also very high along with the ASR value, which means that in this case, our attack worked perfectly because it is unlikely to be detected by the server, and our attack can act as a ghost in the system. As we will discuss only the successful cases in this paragraph, the next case that was able to cross the threshold values was the poisoning of 4\% of the feature column values. After training, we tested the accuracy of the server, which was determined as 0.8611. In addition, the ASR for this 4\% poisoning case was also transcribed at 0.8616, which is a clear indication of a successful attack scenario, because both values were higher than the threshold. Subsequently, a successful case was observed with 20\% values of poisoned data in the feature column, where the accuracy of the server was 0.7427, and the ASR value was reported as 0.7763. The next successful case was with 25\% of poisoned values in the feature column of the dataset, and after training when the server was tested, the accuracy reached 0.9680, while the ASR also increased to 0.9671.

We discuss some unsuccessful cases in $N_{BAU1}^{CIC-FP}$ scenario, where the first case is transcribed with 2\% and 3\% of the poisoned data. With 2\% feature-poisoned data, the accuracy of the server dropped to 0.2692, and the ASR value also exhibited a significant decline to 0.2809. Both values are less than the threshold; therefore, they fall under the unsuccessful cases. Additionally, with a 3\% FP attack, the server accuracy decreased to 0.0491 and the ASR also decreased to 0.0575, which again fulfills the requirements of the unsuccessful case. Next, such cases were observed with 5\%, 7\%, 10\%, and 15\% of the feature-poisoned data, where the accuracy in all cases was less than 0.3500, which is less than the threshold value. Furthermore, the ASR value in these three cases was less than 0.36, which is again less than our threshold. These attack cases cannot fool the server and are easily detected by the system.  
\begin{table}[h!]
    \centering
    \renewcommand{\arraystretch}{1.3}
    \caption{Feature Poison (FP) Attack for the Scenario $N_{BAU1}^{CIC-FP}$ on CIC Dataset}
    \begin{tabular}{|c|c|c|c|c|} \hline 
         Poison& Component &Loss & Accuracy & ASR\\ \hline 
         1\%&Client-1 \& Client-2 & 0.6507 \& 0.6507 & - & -  \\ \hline 
         \rowcolor{gray!30}
         1\%& \textbf{Server} & - &0.9642 & 0.9628\\\hline
         
         4\% &Client-1 \& Client-2& 0.6754 \& 0.6754 & - & -\\\hline
         \rowcolor{gray!30}
         4\%& \textbf{Server} & - &0.8611 &0.8616\\\hline
        
         20\%&Client-1 \&Client-2 &  0.6785 \& 0.6790 & - & -\\\hline
         \rowcolor{gray!30}
         20\%& \textbf{Server} & - &0.7427 & 0.7763\\\hline
         
         25\%&Client-1 \&Client-2 &  0.6321 \& 0.6321& - & -\\\hline
         \rowcolor{gray!30} 
         25\%& \textbf{Server} & - &0.9680 & 0.9671\\\hline

    \end{tabular}
    \label{tab:with_FP_attack_CIC}
\end{table}

\subsection{Results With Feature Poisoning (FP) Attack on UNSW dataset}
This section is related to scenarios with a feature-poisoning attack on the UNSW dataset. In Table \ref{tab:with_FP_attack_UNSW} we list the successful cases that show the impact of the FP attack on the UNSW dataset. Again, in this scenario, we regarded the poison percentages of the attack as successful, in which the server accuracy and ASR both have values greater than a threshold value, which was set to 0.40. We manipulated the values of one feature of the UNSW dataset and trained the model with 1\% of the manipulated values of the most important features in the UNSW dataset. With 1\% poisoned data, we observed that the accuracy of the server was 0.8195 and, more interestingly, the ASR value was 0.8232. The accuracy of the server is good, and the ASR is also greater than the threshold value, which means that in this case, our attack worked perfectly because it is difficult to detect by the server, and our attack can act as a ghost in the system. The next case that could cross the threshold value was poisoning of the 2\% and 3\% values of the feature column. After training, we tested the accuracy of the server, which was above 0.84 in both cases. Additionally, the ASR for the 2\% and 3\% poisoning cases was also noted to be above 0.81, as shown in Table \ref{tab:with_FP_attack_UNSW} which is a clear indication of successful attack cases because both values are higher than the threshold in both cases. Subsequently, all the poison percentages except 4\% were observed to be successful cases. The accuracy and ASR values at 5\%, 7\%, 10\%, 15\%, 20\%, and 25\%  poisoning percentages are greater than the threshold; therefore, if we poison the UNSW dataset with these percentages, we can fool the system. \\
\indent There was only one unsuccessful case in this scenario in which the poison percentage was set to 4\%, and when the model was trained and tested after manipulating the values according to this percentage of the feature column, we obtained a server accuracy value of 0.0952, while the ASR value was also observed to be 0.1089. 
\begin{table}[h!]
    \centering
    \renewcommand{\arraystretch}{1.3}
    \caption{Feature Poison Attack for the Scenario $N{BAU1}^{UNSW-FP}$ on UNSW Dataset}
    \begin{tabular}{|c|c|c|c|c|c|} \hline 
         Poison & Component & Loss & Acc & ASR\\ \hline 
         1\% & Client-1 \& Client-2 & 0.6779 \& 0.6780 & - & -  \\ \hline 
         \rowcolor{gray!30}
         1\% & \textbf{Server} & - & 0.8195 & 0.8231\\ \hline
         
         2\% & Client-1 \& Client-2 & 0.6769 \& 0.677  & - & -\\ \hline
         \rowcolor{gray!30}
         2\% & \textbf{Server} & - & 0.8527 & 0.8722\\ \hline
         
         3\% & Client-1 \& Client-2& 0.672 \& 0.672 & - & -\\ \hline
         \rowcolor{gray!30}
         3\% & \textbf{Server} & - & 0.8433 & 0.8194\\ \hline
         
         5\% & Client-1 \& Client-2 & 0.6729 \&  0.6728 & - & -\\ \hline
         \rowcolor{gray!30}
         5\% & \textbf{Server} & - & 0.8620 & 0.8491\\ \hline
         
         7\% & Client-1 \& Client-2 & 0.6691 \& 0.6690 & - & -\\ \hline
         \rowcolor{gray!30}
         7\% & \textbf{Server} & - & 0.9017 & 0.9009\\ \hline
         
         10\% & Client-1 \& Client-2 & 0.6718 \& 0.6723& - & - \\ \hline
         \rowcolor{gray!30}
         10\% & \textbf{Server} & - & 0.8725 & 0.8944\\ \hline
         
         15\% & Client-1 \& Client-2 & 0.6454 \& 0.6462 & - & -\\ \hline
         \rowcolor{gray!30}
         15\% & \textbf{Server} & - & 0.9066 & 0.9086\\ \hline
         
         20\% & Client-1 \& Client-2 & 0.6939 \&  0.6939 & - & -\\ \hline
         \rowcolor{gray!30}
         20\% & \textbf{Server} & - & 0.4682 & 0.4824\\ \hline
         
         25\% & Client-1 \& Client-2 & 0.6640 \&0.6649 & - & -\\ \hline
         \rowcolor{gray!30}
         25\% & \textbf{Server} & - & 0.9018 & 0.9064\\ \hline
        \end{tabular}
    \label{tab:with_FP_attack_UNSW}
\end{table}

\section{Conclusions and Future works}
\label{Conclusions}
In this study, we explored the vulnerability of FL models to data poisoning attacks in the computer network domain. For this purpose, we carried out two data poisoning attacks, LF and FP, on two prominent datasets, CIC and UNSW, which are part of numerous studies in computer networks. We divided the experiments into four different scenarios for both attacks and both datasets, and reported the results in different tables in Section \ref{Results}. Our findings provide critical insights into the susceptibility of FL models to adversarial attack. We determined that the LF attack is not an effective approach to fool the server because the accuracy drops significantly, and the system can notice it immediately and raise a critical alert. However, the FP attack showed significantly more effective results and was able to fool the server in most cases in both scenarios. Feature importance analysis revealed that a few features in the dataset can significantly influence the decision-making process of the model. If we successfully find these important features and exploit them, then these vital features in FP attacks require enhanced feature-protection mechanisms.

In future work, enhanced advanced defense strategies can be proposed to defend the FL against data-poisoning attacks. Studies can focus on feature obfuscation and protection strategies that can reduce the influence of feature-poisoning attacks, including feature-level encryption and feature-perturbation techniques. Moreover, researchers should delve into more sophisticated adversarial strategies, including transfer attacks between models, evasion attacks, and adversarial federated learning, to bolster defense against these evolving threats. By identifying attack vectors and their critical impacts, we can establish a framework for developing more resilient and enduring FL systems and enhancing their capability to face challenging threats in a dynamic cybersecurity environment. In addition, we experimented with two clients; however, in future studies, it can be extended to N clients, where the value of N is greater than 2, and more interesting results can be expected.


%

\bibliographystyle{IEEEtran}
\bibliography{References}

\vskip -3\baselineskip plus -1fil
\begin{IEEEbiography}
[{\includegraphics[width=1in,height=1.25in]{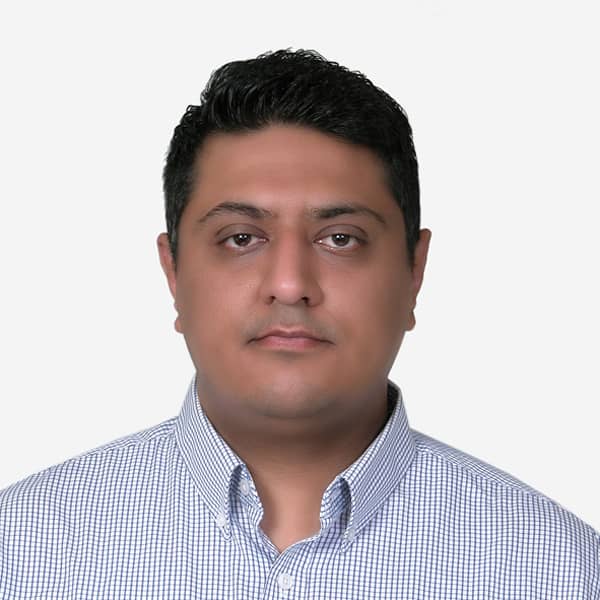}}]
{Ehsan Nowroozi} (Senior Member, IEEE) is a Senior Lecturer (Associate Professor) at Ravensbourne University London, Department of Business and Computing, London, UK. He received his doctorate from Siena University in 2020. His research has had a significant impact on the development of AI-cyber and the ability to defend against cyber threats. He had four postdocs in different high-prestige universities, including a Research Fellow at the Centre for Secure Information Technologies (CSIT) at Queen’s University Belfast in the United Kingdom, a Research Fellow at the Security and Privacy Research Group (SPRITZ) at the University of Padua in Italy, a Research Fellow at the Visual Information Processing and Protection (VIPP) at the University of Siena in Italy, a Research Fellow at the Sabanci University in Turkey. He was also an assistant professor at Bahcesehir University, Istanbul, Turkey. He has worked on a variety of projects funded by renowned institutions, such as DARPA, the Air Force Research Laboratory (AFRL) of the U.S. government, the Italian Ministry of University and Research (MUR), and THALES United Kingdom. He serves as a reviewer for prominent journals, such as IEEE TNSM, IEEE TIFS, and IEEE TNNLS. In addition, he has been a senior member of the Institute of Electrical and Electronics Engineers (IEEE) since 2022 and an ACM Professional member since 2023.
\end{IEEEbiography}

\vskip -3\baselineskip plus -1fil
\begin{IEEEbiography}[{\includegraphics[width=1in,height=1.25in]{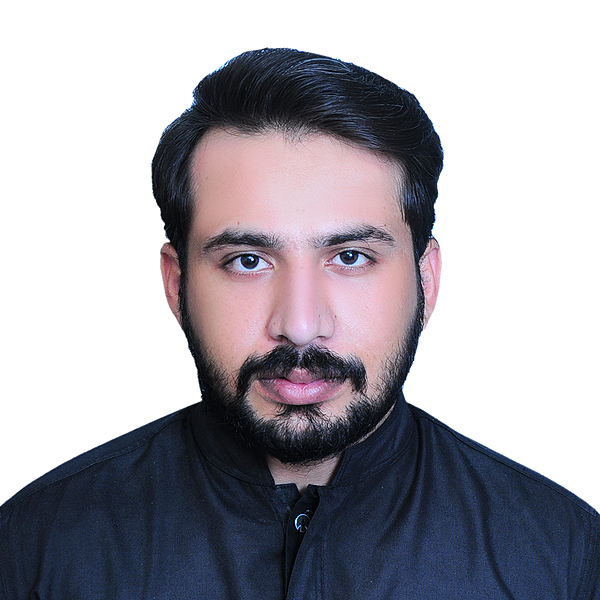}}]{Imran Haider} (Member, IEEE) Imran is a cybersecurity master's student at Bahcesehir University in Istanbul, Turkey. He completed his bachelor’s degree in computer science from the National University of Computer and Emerging Sciences (NUCES), Pakistan. Imran is a student member of IEEE and possesses a strong passion for Cybersecurity and Artificial Intelligence research. In addition to his two years of experience in software development and proficiency in Python for machine/deep learning. He also holds expertise in penetration testing—an essential skill for identifying and mitigating security vulnerabilities in digital systems.  
\end{IEEEbiography}

\vskip -2\baselineskip plus -1fil
\begin{IEEEbiography}[{\includegraphics[width=1in,height=1.25in]{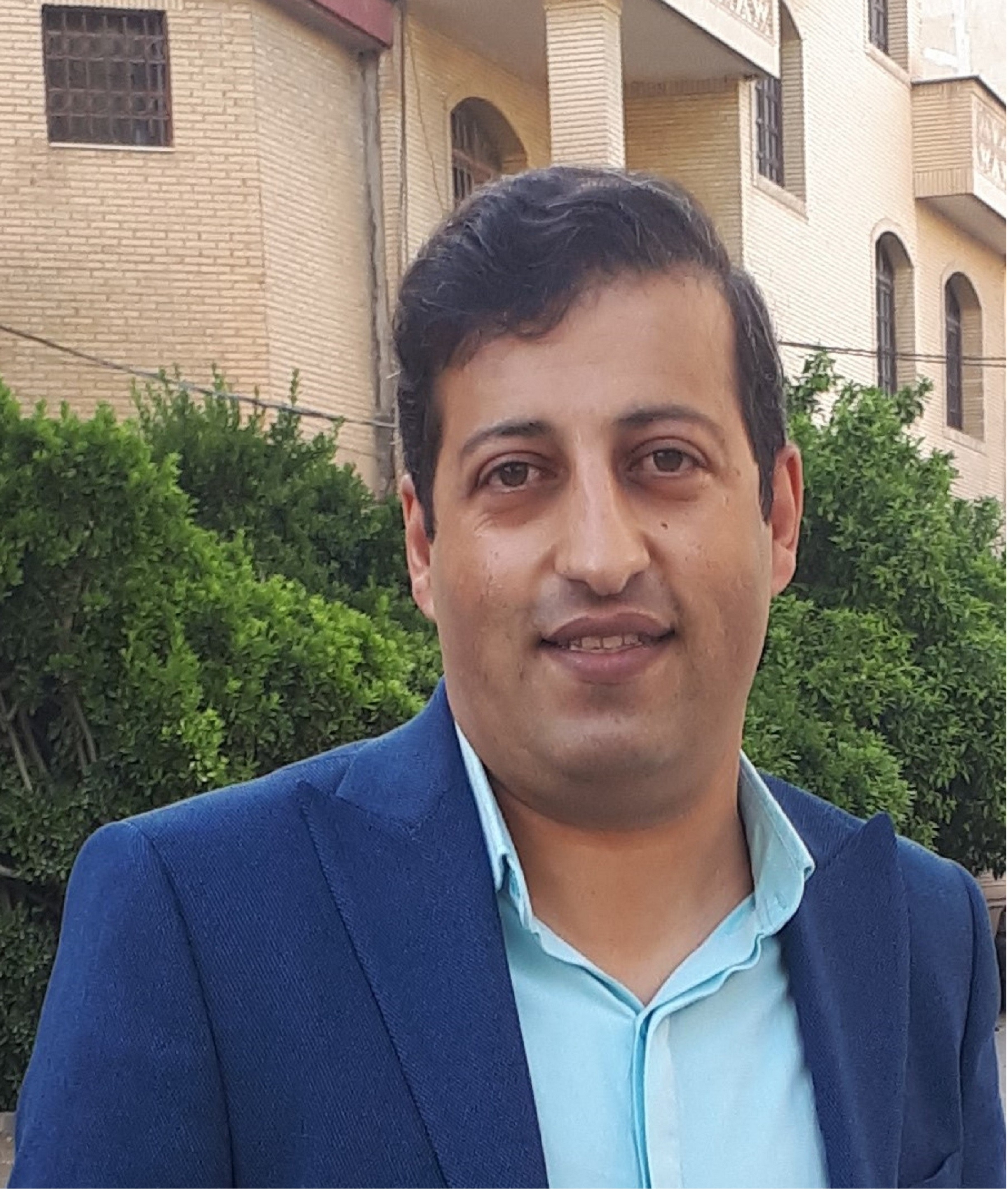}}]{Rahim~Taheri} (Member, IEEE) received his Ph.D. degree in Information Technology from Shiraz University of Technology, Iran, in 2020. Now he is a Lecturer in Cyber Security and Forensics at the University of Portsmouth, UK. Before joining the University of Portsmouth, he was a post-doctoral research associate at King’s Communications, Learning, and Information Processing (Kclip) Lab, King’s College London, UK. His main research interests include machine learning applications in security, adversarial machine learning, and federated learning.
\end{IEEEbiography}

\vskip -2\baselineskip plus -1fil
\begin{IEEEbiography}[{\includegraphics[width=1in,height=1.25in]{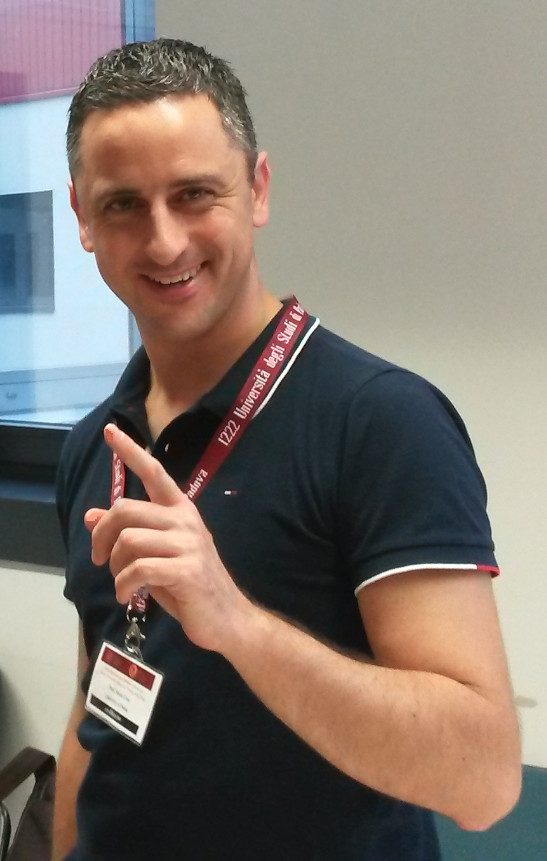}}]{Mauro Conti} (Fellow, IEEE) received the Ph.D. degree from the Sapienza University of Rome, Italy, in 2009. He is a Full Professor at the University of Padua, Italy. He is also affiliated with TU Delft and the University of Washington, Seattle. His research in the area of Security and Privacy is also funded by companies, including Cisco, Intel, and Huawei. He published more than 500 papers in topmost international peer-reviewed journals and conferences. He
is the Editor-in-Chief for IEEE TRANSACTIONS ON INFORMATION FORENSICS AND SECURITY
and has been an Associate Editor for several journals, including IEEE COMMUNICATIONS SURVEYS AND TUTORIALS, IEEE TRANSACTIONS ON DEPENDABLE AND SECURE COMPUTING, IEEE TRANSACTIONS ON NETWORK AND SERVICE MANAGEMENT. He is a Fellow of YAE and a Senior Member of ACM.

\end{IEEEbiography}

\end{document}